\pdfoutput=1
\documentclass[twocolumn,english,aps,superscriptaddress,prb]{revtex4-1}

\usepackage{babel}
\usepackage{amsmath}
\usepackage{amssymb}
\usepackage{graphicx}

\begin{document}

\title{Lifshitz point at commensurate melting of 1D Rydberg atoms}

\author{Natalia Chepiga}
%%\affiliation{Institute for Theoretical Physics, University of Amsterdam, Science Park 904 Postbus 94485, 1090 GL Amsterdam, The Netherlands}
\affiliation{Department of Quantum Nanoscience, Kavli Institute of Nanoscience, Delft University of Technology, Lorentzweg 1, 2628 CJ Delft, The Netherlands}
 \author{Fr\'ed\'eric Mila}
 \affiliation{Institute of Physics, Ecole Polytechnique F\'ed\'erale de Lausanne (EPFL), CH-1015 Lausanne, Switzerland}

\date{\today}
\begin{abstract} 
The recent investigation of chains of Rydberg atoms has brought back the problem of commensurate-incommensurate transitions into the focus of current research. In 2D classical systems, or in 1D quantum systems, the commensurate melting of a period-p phase with p larger than 4 is known to take place through an intermediate floating phase where correlations between domain walls or particles decay only as a power law, but when p is equal to 3 or 4, it has been argued by
Huse and Fisher that the transition could also be direct and continuous in a non-conformal chiral universality class with a dynamical exponent larger than 1. This is only possible however if the floating phase terminates at a Lifshitz point before reaching the conformal point, a possibility debated since then. Here we argue that this is a generic feature of models where the number of particles is not conserved because the exponent of the floating phase changes along the Pokrovsky-Talapov transition and can thus reach the value at which the floating phase becomes unstable. Furthermore, we show numerically that this scenario is realized in an effective model of the period-3 phase of Rydberg chains in which hard-core bosons are created and annihilated three by three: The Luttinger liquid parameter reaches the critical value $p^2/8=9/8$ along the Pokrovsky-Talapov transition, leading to a Lifshitz point that separates the floating phase from a chiral transition. Implications beyond Rydberg atoms are briefly discussed.
\end{abstract}

\maketitle

%%%%%%%%%%%%%%%%%%%%%%%%%%%%%%%%%%%% INTRODUCTION %%%%%%%%%%%%%%%%%%%%%%%%%%%%%%%%%%%%

\section{Introduction}

Rydberg atoms trapped with optical tweezers are becoming one of the major playgrounds to investigate quantum matter. The laser detuning, which plays the role of the chemical potential and controls the number of excited atoms, can be easily tuned, and its interplay with the long-range van der Waals repulsion and the creation and annihilation of excited states by a laser with appropriate Rabi frequency has opened the way to a full experimental mapping of the phase diagram in one-dimension\cite{Bernien2017,kibble_zureck}. This phase diagram is dominated at large detuning by big lobes of density-waves of simple integer periods\cite{Bernien2017,rader2019floating,kibble_zureck}, and at  small detuning by a disordered phase with short-range, incommensurate correlations\cite{prl_chepiga,chepiga2020kibblezurek}. What happens between these main phases is remarkably rich however. At very large detuning, devil's staircases of incommensurate density waves\cite{Bak_1982} are expected to be present because of the long-range character of the repulsion between atoms\cite{rader2019floating}. But even at intermediate detuning, where these phases are not present, the transition between the integer-period density waves and the disordered incommensurate phase is a very subtle issue\cite{fendley,prl_chepiga,chepiga2020kibblezurek,giudici,samajdar}. The problem is the 1D quantum analog of the famous problem of commensurate-incommensurate (C-IC) transitions in classical 2D physics\cite{Ostlund,huse,HuseFisher,Selke1982,HUSE1983363,HuseFisher1984,Duxbury,yeomans1985,HOWES1983169,howes1983,houlrik1986,bartelt,Den_Nijs,Everts_1989,birgeneau,SelkeExperiment,sato,PhysRevB.92.035154}, a problem still partially unsolved in spite of four decades of analytical and numerical work. With their tunability, Rydberg atoms open new perspectives in the experimental investigation of the entire boundary of these transitions, and in the possible resolution of the puzzles still posed by the C-IC transition.

A priori, one could expect the transition out of period-$p$ phases to be simply in a universality class controlled by the value of $p$ (Ising for $p$=2, 3-state Potts for $p$=3, Ashkin-Teller for $p$=4, etc.). However, when the transition is driven by the proliferation of domain walls between ordered domains simply related to each other by translation, the disordered phase is incommensurate. The asymmetry between domain walls induces a chiral perturbation\cite{HuseFisher} that is in most cases relevant and has to drive the transition away from the standard universality classes. For $p\ge 5$, it is well accepted that the transition becomes a Pokrosky-Talapov\cite{Pokrovsky_Talapov,schulz1980} transition into a critical phase, followed by a Kosterlitz-Thouless\cite{Kosterlitz_Thouless} transition into a disordered phase with exponentially decaying correlations\cite{Den_Nijs}. By analogy to the classical 2D problem, the intermediate critical phase is referred to as the floating phase. 

For $p=3$ and $p=4$, there is no consensus however. If, in the disordered phase, there is a line where the short-range correlations have the periodicity of the adjacent ordered phase, as is the case for Rydberg chains, the transition is expected to be a standard transition in the 3-state Potts and Ashkin-Teller universality classes respectively\cite{HuseFisher1984}. This has been explicitly demonstrated for models with infinite short-range repulsions for period-3 \cite{fendley,prl_chepiga} and period-4 phases\cite{chepiga2020kibblezurek}. Far from this commensurate line, it is also by now fairly well established numerically that, as for $p\ge 5$, there is a floating phase\cite{rader2019floating}. The main issue is what happens in between. In 1982, Huse and Fisher\cite{HuseFisher} argued that the floating phase may not appear immediately, and that the transition, which cannot be in a standard universality class because the chiral perturbation is relevant, could still be direct and continuous, but in a new universality class that they called chiral. The presence of such a chiral transition is consistent with the interpretation of recent Kibble-Zurek experiments on Rydberg chains\cite{chepiga2020kibblezurek,kibble_zureck}. However, it has not been possible so far to come up with a compelling theoretical argument in favour of a Lifschitz point that would terminate the floating phase at a distance from the Potts point for $p=3$ or the Ashkin-Teller point for $p=4$, and in the absence of such an argument, the issue remains controversial.

In the present paper, we come up with such an argument. We show that the instability of the floating phase is driven by a property of the Pokrovsky-Talapov transition that has apparently been overlooked so far, namely that the Luttinger liquid exponent of the floating phase can change along this transition. Then, if the Luttinger liquid exponent reaches the value at which the floating phase becomes instable, which is the mechanism behind the Kosterlitz-Thouless transition from the floating phase into the disordered one, the transition can no longer take place through an intermediate phase, opening the way to a chiral transition. We further argue that the Pokrovsky-Talapov transition is expected to have this property for models where the number of particles is not conserved, and we prove it in the case of a 1D model where particles are created and annihilated three by three, a model put forward recently in the context of Rydberg atoms\cite{PhysRevB.98.205118}, but already introduced earlier in the fermionic description of domain walls  in the context of the C-IC transition in 2D classical systems\cite{Den_Nijs}.

\section{General argument}

\begin{figure}[t!]
\centering 
\includegraphics[width=0.45\textwidth]{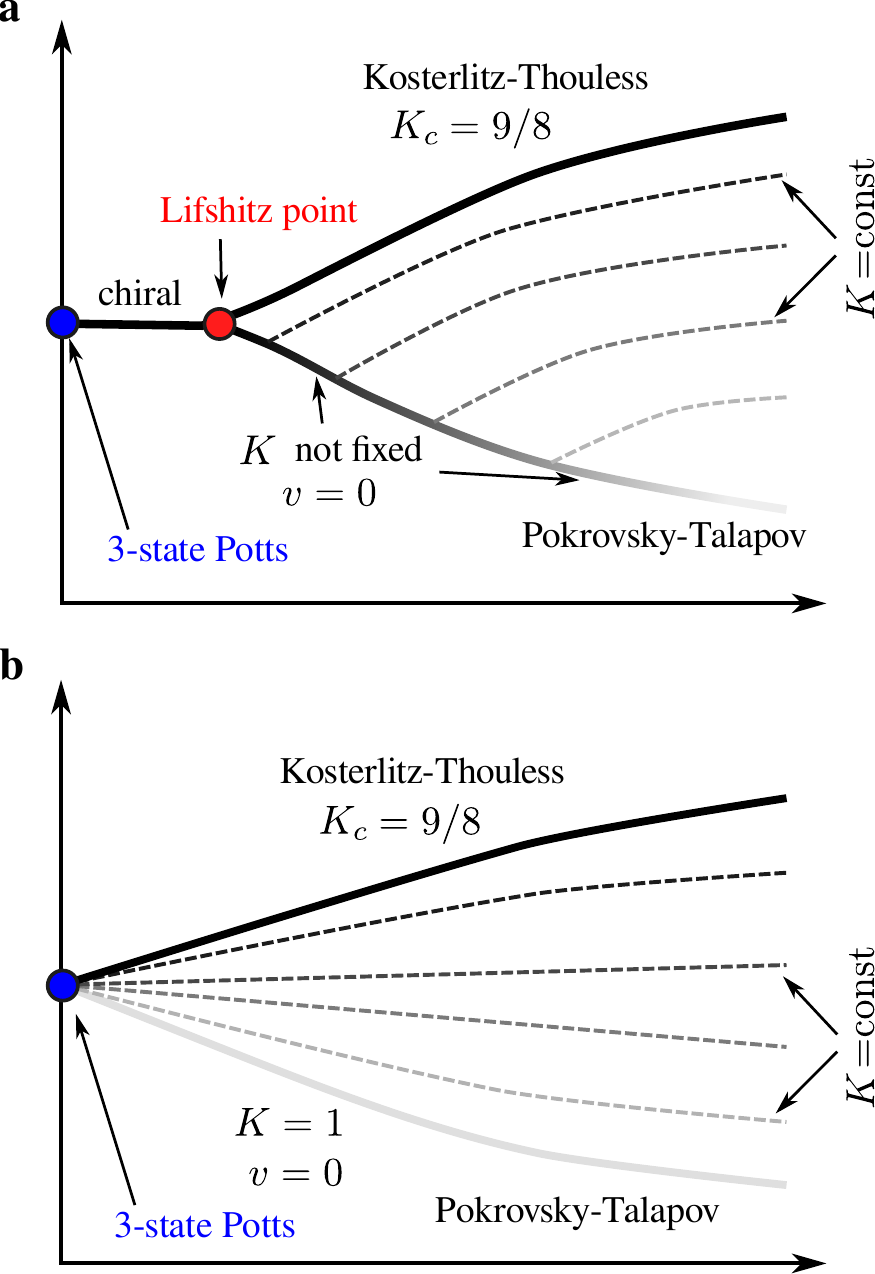}
\caption{(Color online) Sketches of the two main possibilities for the phase diagram of commensurate melting. {\bf a} In the absence of chiral perturbation the transition is conformal in the three-state Potts universality class (blue point). If the Pokrosvky-Talapov transition does not lead to an empty phase, the Luttinger liquid exponent changes along the Pokrovsky-Talapov transition. It reaches the value $K=K_c$ of the Kosterlitz-Thouless transition at the Lifshitz point (red dot), beyond which the transition is chiral in the Huse-Fisher universality class. {\bf b} When the phase below the Pokrosvky-Talapov transition is empty, the Luttinger liquid exponent tends to the non-interacting value $K=1$ when approaching this line. All constant $K$ lines meet at the three-state Potts critical point.}
\label{fig:TheorySketches}
\end{figure}

From now on, we will concentrate on the case $p=3$ for clarity. The argument can be straightforwardly extended to $p=4$. Let us assume that there is a line where the correlations remain commensurate in the disordered phase along which the transition is in the 3-state Potts universality class, and that there is a floating phase further away along the transition. Then either the floating phase starts right away, as in the bottom panel of Fig.\ref{fig:TheorySketches}, or it only starts at a Lifshitz point different from the Potts point, as in the top panel of Fig.\ref{fig:TheorySketches}.

In the language of 1D quantum physics, the floating phase is a Luttinger liquid\cite{giamarchi}, and it is described by two parameters: i) The parameter $K$ that controls the decay of all correlation functions, often referred to as the Luttinger liquid exponent; ii) The velocity $v$ that controls the small momentum dispersion of the excitations. At a C-IC transition, this intermediate phase is bounded by two very different transitions, and each of them is a priori controlled by a single parameter: 

- The parameter $K$ controls the Kosterlitz-Thouless (KT) transition into the disordered phase with exponentially decaying correlations. This transition occurs when an operator present in the model (or generated under the renormalization group flow) becomes relevant, i.e. when its scaling dimension becomes smaller than 2. For the operator that simultaneously creates $p$ domain walls or particles, this scaling dimension is equal to $p^2/4K$ in a Luttinger liquid with exponent $K$\cite{sachdev_QPT}, and this operator becomes relevant for $K>K_c$ with $K_c=p^2/8$.

- The parameter $v$ controls the Pokrosky-Talapov transition. At this transition, $v$ goes to zero. The dispersion becomes quadratic, and the dynamical exponent is equal to 2. 

There is a priori no reason for the Luttinger liquid parameter to have a specific value along the Pokrovsky-Talapov (PT) transition since this transition is controlled by the velocity. However, it is a well known fact that the Luttinger liquid exponent is often constant along transition lines\cite{cazalilla,giamarchi}, and this also applies to the Pokrovsky-Talapov transition in certain cases. In particular, the PT transition describes the transition at which a fermionic system starts to fill up. If the Hamiltonian conserves the number of fermions, the system is empty on one side of the transition. Then, on the other side, in the Luttinger liquid phase, the density goes continuously to zero at the transition. In that case, and for interactions that decay fast enough, the Luttinger liquid exponent is expected to tend to the non-interacting value $K=1$ at the transition. Then the floating phase is limited on one side by $K=K_c$, which is larger than 1 as soon as $p\ge 3$, and on the other side by $K=1$. If the two lines merge at a point along the transition, this point should correspond to the point where all constant $K$ lines with $1\le K \le K_c$ meet. This possibility has been discussed by Haldane, Bak and Bohr\cite{haldane_bak} and by Schulz\cite{schulz1983} in the context of the quantum sine-Gordon model, in which case the point at which the lines meet was shown to be in the $p$-state clock universality class by symmetry. This is summarized in the sketch of the bottom panel of Fig.\ref{fig:TheorySketches}, where the point where all constant $K$  lines meets is called 3-state Potts, the standard terminology for $p=3$.

However, if the density of particles does not go to zero at the Pokrovsky-Talapov transition, and this will in particular be the case if the number of particles is not conserved,  the Luttinger liquid exponent is not fixed at the non-interacting value $K=1$ along this transition. Then, the constant $K$ lines do not have to meet at a single point, but they can terminate at different points along the PT transition, as sketched in the top panel of Fig.\ref{fig:TheorySketches}. This opens the possibility of a Lifshitz point defined as the point where the line $K=K_c$ hits the PT transition. This point is not a point of special symmetry, and there is no reason for this point to be $p$-state clock. If the Lifshitz point occurs before the Potts point, then between them the transition must be the chiral transition predicted by Huse and Fisher\cite{HuseFisher}. 

\section{Model with 3-site term}

We will now show that this is precisely what happens in a hard-core boson model recently proposed as a dual description of the period-$p=3$ transition of 1D Rydberg atoms\cite{PhysRevB.98.205118}. This model is defined by the Hamiltonian:
%Understanding  the  nature  of  quantum  phase  transitions  in  low-dimensional  systems  is  one  of  the  central topics in condensed matter physics. Over past decades a lot of attention has been attracted by critical theories in the presence of chiral perturbations. One of the well-known examples is a commensurate-incommensurate transition out of period-3 phase that has a long history in the context of adsorbed monolayers, chiral clock model and quantum chains of Rydberg atoms. In each of these models translation symmetry is spontaneously broken in the ordered phase. A model that realize Z$_3$ ordered phase without breaking translation symmetry has been proposed recently in the Ref.\onlinecite{PhysRevB.98.205118} and is given by the following quantum many-body Hamiltonian:
\begin{equation}
  H=\sum_i-t(d^\dagger_id_{i+1}+\mathrm{h.c.})-\mu n_i+\lambda(d^\dagger_id^\dagger_{i+1}d^\dagger_{i+2}+\mathrm{h.c.}),
  \label{eq:hamilt}
\end{equation}
Without loss of generality we will fix $t=1$ in the following. 

\begin{figure}[t!]
\centering 
\includegraphics[width=0.45\textwidth]{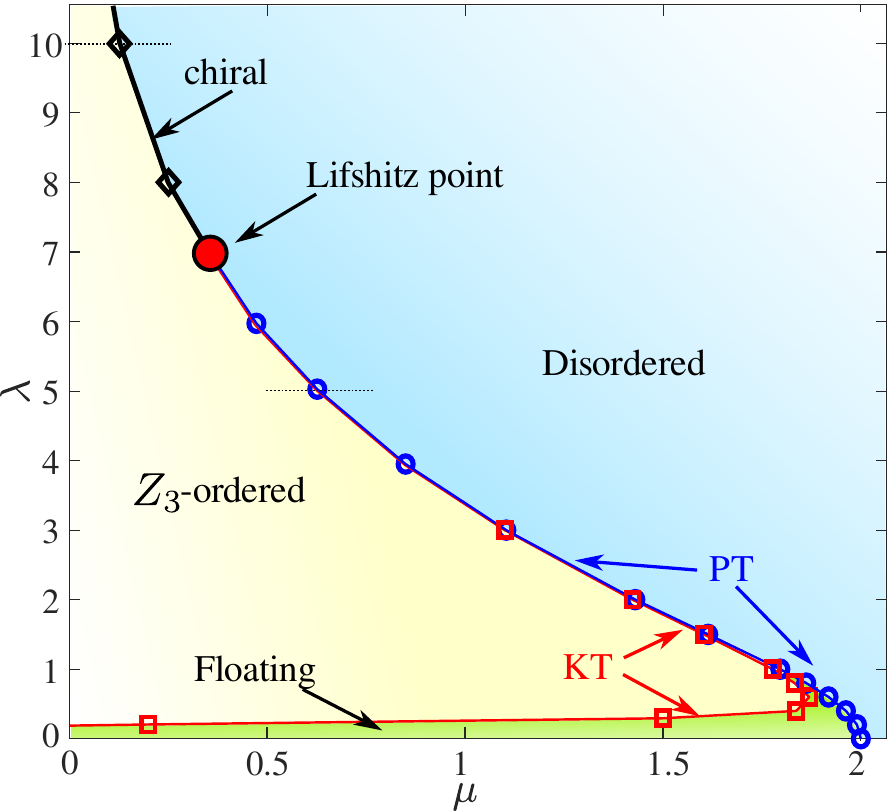}
\caption{(Color online) Phase diagram of the hard-core boson model with three-site term of Eq.\ref{eq:hamilt}. It consists of three phases: (i) A disordered phase (blue) at large $\mu$; (ii) A Z$_3$ ordered phase (yellow) at small $\mu$ and not too small $\lambda$; (iii) A floating phase (green) that starts at small values of $\lambda$ for small $\mu$, and that extends up to $\lambda\approx 7$ upon approaching the disordered phase. The floating phase is separated from the Z$_3$ phase by a Kosterlitz-Thouless transition (red squares), and from the disordered phase by a Pokrovsky-Talapov transition (blue circles). For larger values of $\lambda$ the transition between the disordered phase and the Z$_3$ phase is a direct one in the Huse-Fisher universality class (black diamonds). The dotted lines show the cuts presented in Fig.\ref{fig:cuts}}
\label{fig:phasediag}
\end{figure}

When $\lambda\neq 0$, this model does not conserve the number of particles, and the $U(1)$ symmetry is reduced to Z$_3$. The last term splits the Hilbert space into three sectors distinguished by the total filling modulo 3. The Z$_3$ symmetry is broken if these sectors have the same ground state energy, and unbroken otherwise. We have studied this model numerically with large-scale density matrix renormalization group (DMRG)\cite{dmrg1,dmrg2,dmrg3,dmrg4} simulations on systems with up to 3000 sites keeping up to 2000 states and truncating singular values below $10^{-8}$.
Our numerical results are summarized in the phase diagram of Fig.\ref{fig:phasediag}. The phase diagram is symmetric around $\mu=0$, so we only show and discuss the positive $\mu$ side. There are two gapped phases: i) the disordered phase at large enough $\mu$, which is commensurate with wave-vector zero and corresponds to a full system for $\lambda=0$, and ii) the Z$_3$ ordered phase, with short-range incommensurate correlations. There is also a floating phase, a critical phase in the Luttinger liquid universality class with algebraic incommensurate correlations. Along the vertical line $\mu=0$, the wave-vector vanishes by symmetry, making this line the commensurate line along which the transition should be in the universality class of the 3-state Potts model. As we shall see, the floating phase extends up to a Lifshitz point located at ($\mu\simeq 0.35,\lambda\simeq7$), far from the commensurate line, hence of the 3-state Potts point. Accordingly, beyond this Lifshitz point, the transition must be a direct one in the Huse-Fisher chiral universality class\cite{HuseFisher,HuseFisher1984,PhysRevB.98.205118}. The floating phase is separated from the disordered phase by a Pokrovsky-Talapov transition, and from the Z$_3$ ordered phase by a Kosterlitz-Thouless transition.

Note that the boundary of the disordered phase agrees with the numerical results of Ref.\cite{PhysRevB.98.205118}. In this reference, the authors also reported that the Z$_3$ ordered phase is gapped provided $\lambda$ is large enough, implicitly implying that it might be gapless at small $\lambda$, but they did not try
to determine the boundary where the gap closes. In that respect, our numerical results complement and correct the numerical results of Ref.\cite{PhysRevB.98.205118}. There is indeed a gapless phase at small $\lambda$, but it extends to large values of $\lambda$ in the vicinity of the transition to the disordered phase in the form of very narrow floating phase. The difficulty encountered by the authors of Ref.\cite{PhysRevB.98.205118} to identify the universality class of the transition is probably a consequence of this narrow floating phase. 

Note also that, because of the dual nature of the model, the role of ordered and disordered phases is exchanged with respect to Rydberg atoms. The period-3 phase of Rydberg atoms corresponds to the disordered phase of the model of Eq.\ref{eq:hamilt}, and the disordered phase of Rydberg atoms to  its Z$_3$ ordered phase.

The precise form of this phase diagram has been reached by a careful numerical identification of the various phases and of the transitions between them that we now review.

\begin{figure}[t!]
\centering 
\includegraphics[width=0.45\textwidth]{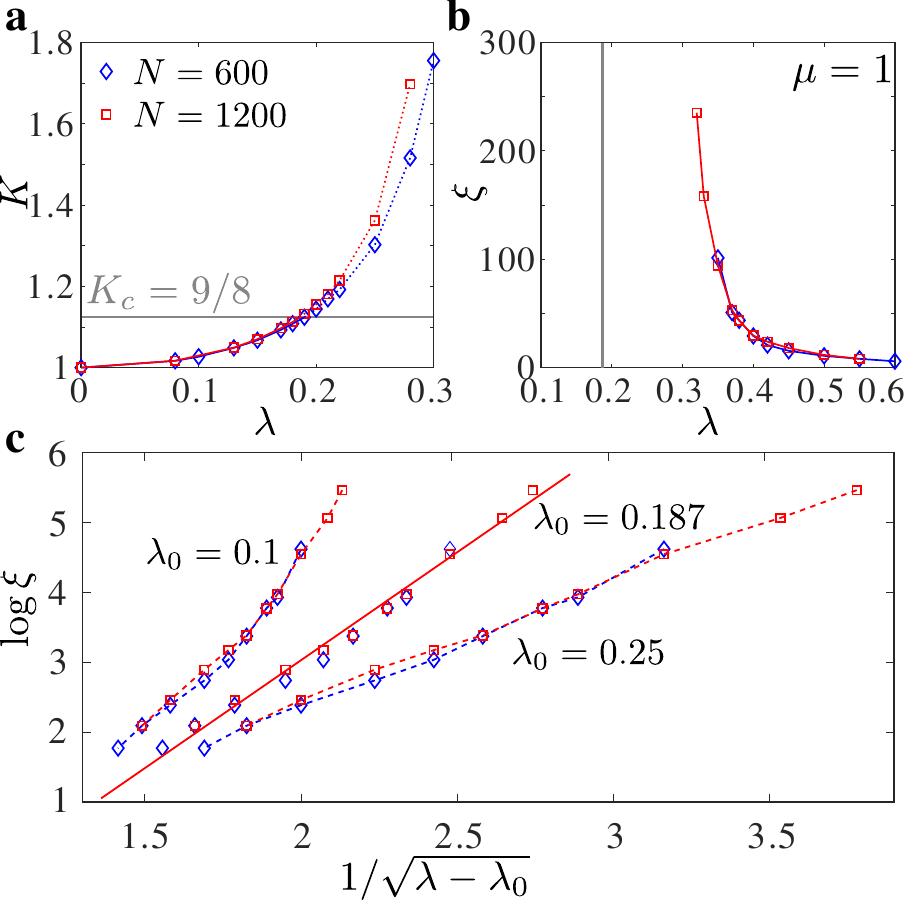}
\caption{Kosterlitz-Thouless transition at $\mu=1$. {\bf a} Luttinger Liquid parameter as a function of $\lambda$. It increases from $K=1$ in the non-interacting case at $\lambda=0$ to the critical value of the Kosterlitz-Thouless transition $K_c=9/8$ at $\lambda_c\approx 0.187$. The large values beyond that point (marked with dashed line) are finite-size effects. The Z$_3$ ordered phase is gapped, and correlations no longer decay as a power law. {\bf b} 
Correlation length in the Z$_3$ ordered phase. It is consistent with a strong divergence upon approaching $\lambda_c$. {\bf c} Scaling of the correlation length in the Z$_3$ ordered phase as a function of $1/\sqrt{\lambda-\lambda_0}$ in a semilog plot. The scaling is linear for $\lambda_0=\lambda_c\approx 0.187$ identified in panel {\bf a}, as expected for a Kosterlitz-Thouless transition, and it is concave for $\lambda_0=0.25$ and convex for $\lambda_0=0.1$, confirming that the transition has to take place at a finite value of $\lambda$. The solid line is a linear fit, dashed lines are guides to the eye.
}
\label{fig:KT}
\end{figure}

\subsubsection{Floating phase and Kosterlitz-Thouless transition}

In the non-interacting case $\lambda=0$,  the model can be mapped on non-interacting spinless fermions. For $\mu<2$, the state is a partially filled band up to a wave-vector $k_F$, and all correlations are critical. Along this non-interacting line, the Luttinger liquid exponent is rigorously equal to $K=1$, including at the Pokrovsky-Talapov commensurate-incommensurate transition to the gapped phase at $\mu=2$.

Upon increasing $\lambda$, the correlations are expected to remain critical as long as all operators acting as perturbations are irrelevant, i.e. have a scaling dimension larger than 2. Now, the operator with the smallest scaling dimension  is expected to be the three boson operator. Indeed, the scaling dimension of an operator creating $m$ particles is equal to $m^2/4K$, and since the number of particles is conserved modulo 3, the only operators allowed by symmetry correspond to creating $m=3n$ fermions with $n$ integer, with an exponent $9n^2/4K$ that is minimal for $n=1$, i.e. for the term creating $p=3$ bosons. Its exponent is equal to $9/4K$, so this operator is {\it irrelevant} along the non-interacting line. It only becomes relevant when $K=9/8$. As a consequence, the critical behaviour of the non-interacting line must extend into a critical floating phase up to the constant-$K$ line $K=9/8$ in the $\lambda-\mu$ phase diagram, a property of the model not discussed in Ref.\onlinecite{PhysRevB.98.205118}.

\begin{figure*}[t!]
\centering 
\includegraphics[width=.98\textwidth]{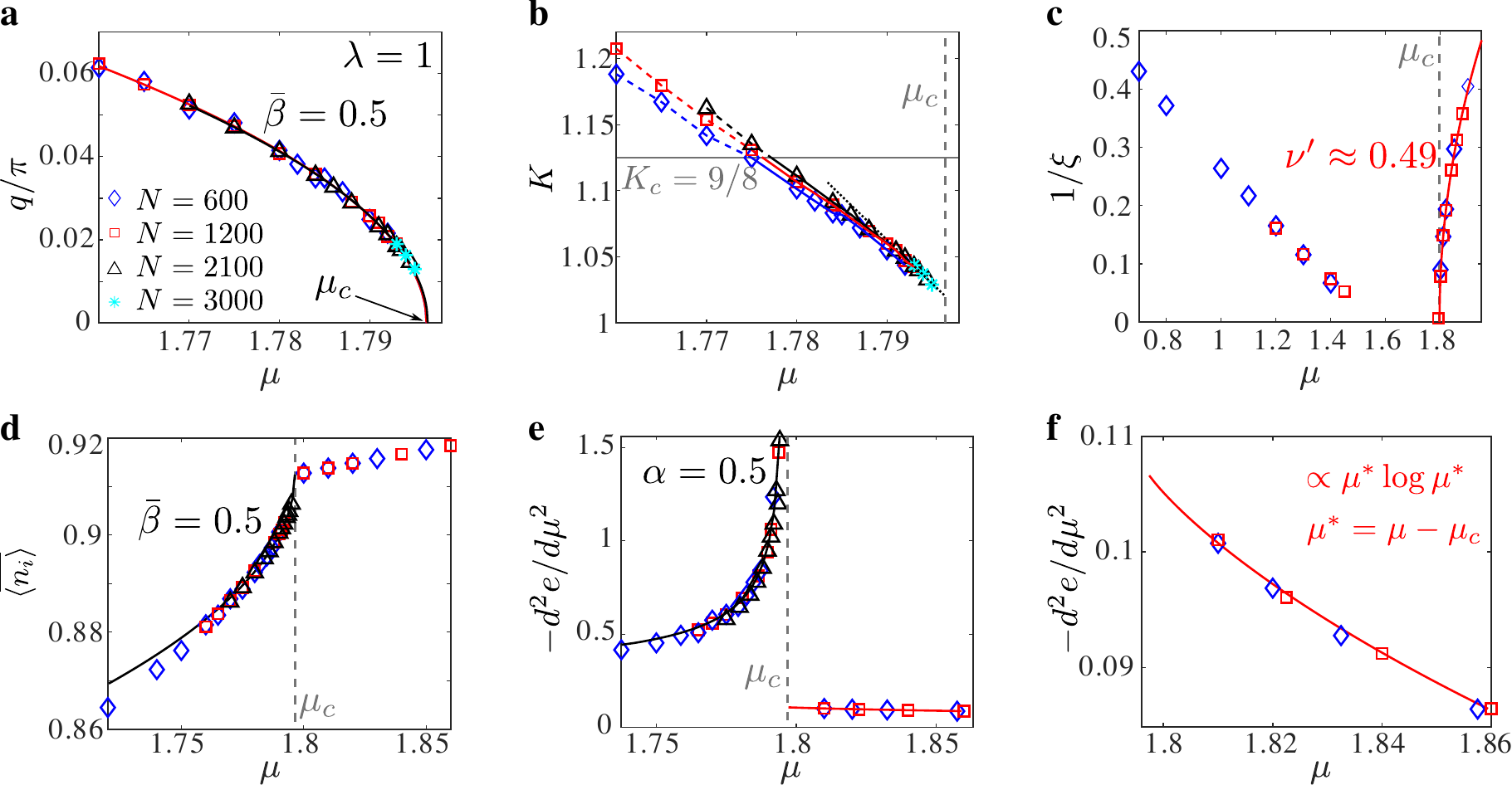}
\caption{Main properties of the Pokrovsky-Talapov transition. {\bf a} Scaling of the incommensurate wave-vector $q$ with the distance to the Pokrovsky-Talapov transition. The solid lines are the results of fits with the Pokrovsky-Talapov critical exponent $\beta=1/2$. The critical point $\mu_c$ identified with these fits is used in panels {\bf b}-{\bf e}. {\bf b} Evolution of the Luttinger liquid parameter $K$. The system is in the floating phase when $K$ is below $K_c=9/8$. Larger values of $K$ (shown as dashed lines) are finite size effects since the Z$_3$ phase is gapped with exponential correlations. The dotted line is a linear extrapolation of the last 3 points. {\bf c} Inverse of the correlation length on both sides on the floating phase. The convex curve on the left is consistent with the exponential divergence expected at the Kosterlitz-Thouless transition. In the disordered phase, the correlation length diverges with the exponent $\bar{\beta}\approx0.49$, consistent with the Pokrovsky-Talapov value $1/2$. {\bf d} Average density across the Pokrovsky-Talapov transition. The solid line is a fit with the Pokrovsky-Talapov critical exponent $\bar{\beta}=1/2$. {\bf e-f} Opposite of the second derivative of energy with respect to $\mu$ - the analog of the specific heat - across the transition. It diverges with critical exponent $\alpha=1/2$ below the transition and approaches a finite value as $(\mu-\mu_c)\log(\mu-\mu_c)$ in the disordered phase.}
\label{fig:BigFig_lambda1}
\end{figure*}

To extract the Luttinger liquid exponent inside the floating phase, we have fitted the Friedel oscillations of the local density profile induced by the open boundary conditions. Details can be found in the Methods section. To demonstrate the validity of the criterion $K=9/8$ for the transition into the disordered phase, we show in Fig.\ref{fig:KT} the evolution of $K$ along a vertical cut $\mu=1$ inside the floating phase, and the correlation length $\xi$ extracted from the density-density correlations in the Z$_3$ ordered phase. Note that since $K$ is expected to change only by $12.5\%$, high accuracy and sufficiently large system sizes are required to detect such changes. At the KT transition, the correlation length is expected to diverge very fast, as $\xi \propto \exp{C/\sqrt{\lambda-\lambda_c}}$. This is consistent with the very steep divergence of $\xi$, and plotting $\ln \xi$ as function of $1/\sqrt{\lambda-\lambda_0}$ for various values of $\lambda_0$ shows that the behaviour is linear at $\lambda_0=\lambda_c$ and concave resp. convex away from it, where $\lambda_c$ is the value at which $K$ reaches the value $K=9/8$. 

The boundary between the floating phase and the Z$_3$ ordered phase is almost horizontal and limited to small values of $\lambda$ up to $\mu \simeq 1.8$, but then it turns up and slowly approaches the boundary to the disordered phase.  For $\lambda>1$, the floating phase is extremely narrow, with a width $\Delta\mu <0.02$.  The key qualitative question is whether the two lines get asymptotically close as $\lambda \rightarrow +\infty$, or whether they meet at a finite value of $\lambda$, which would signal the presence of a Lifshitz point. To address this question, we now turn to a careful investigation of the transition between the floating phase and the disordered phase.

\subsubsection{Pokrovsky-Talapov transition}

For $\lambda=0$, the transition at $\mu=2$ is just a transition between a completely filled band for $\mu>2$ and a partially filled band for $-2<\mu<2$ in terms of spinless fermions. The density is equal to 1 for $\mu>2$ and decreases with a singularity $(2-\mu)^{1/2}$ for $\mu<2$. In the partially filled band, all correlation functions decay as power laws with an oscillating prefactor $\cos(k r)$, where $k$ is a multiple of the Fermi wave-vector $k_F$. In particular, the density-density correlations decay as $\cos(q r)/r^2$, with $q=2k_F$. Since between $\mu=-2$ and $\mu=2$ the Fermi wave-vector grows continuously from $0$ to $\pi$, the wave-vector $q$ is generically incommensurate. Close to $\mu=2$, $q$ approaches 0 modulo $2\pi$ as $(2-\mu)^{1/2}$. At the transition point itself the velocity vanishes and the dispersion is quadratic, so that the dynamical exponent is given by $z=2$. For $-2 < \mu <2$, the Luttinger liquid exponent is fixed a $K=1$. Finally, the second derivative of the energy, the equivalent of the specific heat for quantum transition, vanishes identically above $\mu=2$ and diverges with exponent $\alpha=1/2$ when $\mu\rightarrow 2$ from below.

This transition is a special example of a Pokrosky-Talapov transition, and most of its characteristics are generic to this universality class, but not all. So let us review in detail the general properties of the Pokrovsky-Talapov universality class. This is a very asymmetric transition with a dynamical exponent $z=2$. On the commensurate side (the empty or full side for free fermions), the correlation length diverges with an exponent $\nu=1/2$, the specific heat goes to a constant with a cusped singularity due to a logarithmic correction, as shown by Huse and Fisher\cite{HuseFisher1984}, and the density is in general not constant but approaches its value at the critical point without any power-law singularity. On the incommensurate side, the system is described by a Luttinger liquid with a velocity that vanishes at the transition and with an exponent $K$ that can take a priori any value. The wave-vector of the correlations is expected to go to the commensurate one as $|\mu_c-\mu|^{1/2}$, and the density increases or decreases from its value at $\mu_c$ with a singularity $(\mu_c-\mu)^{1/2}$. 
Finally, as discussed above, the Luttinger liquid exponent is not fixed by symmetry at the PT transition. 

This Pokrovsky-Talapov universality class is expected to be realized upon reducing $\mu$ from the disordered commensurate phase if the transition leads to an intermediate floating phase with critical correlations. Since the behaviour along the PT transition is central to our analysis, let us first carefully check the properties of this transition for $\lambda >0$ but not too large. Our numerical results for a horizontal cut at $\lambda=1$ are summarized in Fig.\ref{fig:BigFig_lambda1}. All the properties expected for a PT transition are realized to a high degree of accuracy. We extract the location of the PT transition by fitting the values of $q$ as a function of chemical potential $\mu$ to the form $(\mu_c-\mu)^{\bar{\beta}}$ with the critical exponent $\bar{\beta}=1/2$ as shown in Fig.\ref{fig:BigFig_lambda1}(a). As one can see, finite-size effects for the system sizes shown are already negligible. Considering the other characteristics, the transition is clearly very asymmetric, and all expected critical exponents are consistent with our numerical results. The density is already significantly smaller than 1 at the transition, and it decreases with a square root singularity upon entering the floating phase. Since the density is not fixed to 1 at the transition, the system is not empty in hole language, and the Luttinger liquid exponent is not fixed to 1. And indeed, although it decreases upon approaching the PT transition, it is consistent with a value definitely larger than 1 at the transition. 

\begin{figure}[t!]
\centering 
\includegraphics[width=0.45\textwidth]{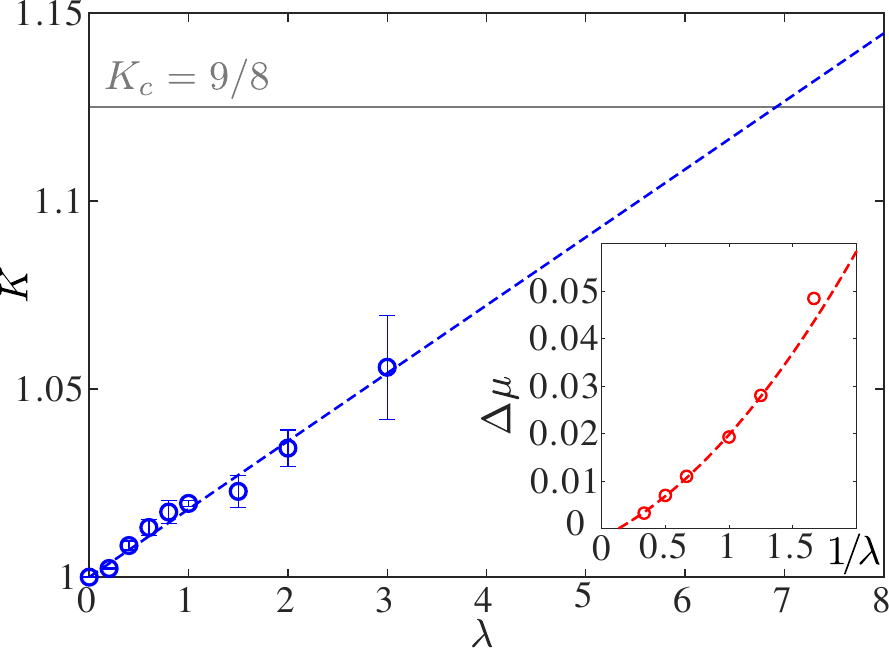}
\caption{(Color online) Luttinger liquid critical exponent $K$ as a function of $\lambda$ along the Pokrovsky-Talapov critical line. Error bars are estimated by extrapolating $K$ as shown in Fig.~\ref{fig:LLatPT}(c),(d) over different subsets of data points and for different system sizes. The dashed line is a linear extrapolation. It reaches the value $K=9/8$ at $\lambda\approx7$. Inset: Width $\Delta \mu$ of the floating phase as a function of $1/\lambda$.}
\label{fig:Kincrease}
\end{figure}

\subsubsection{Lifshitz point and chiral transition}

To extract a quantitative estimate of the Luttinger liquid exponent at the transition point, we have extrapolated the last 3-4 points with linear fits, and we have performed the extrapolation over different sets of points and for various system sizes to estimate the error bars. The evolution of the Luttinger liquid exponent extracted in this way along the PT transition is shown in Fig.\ref{fig:Kincrease}. This is the central numerical result of this paper: In agreement with our hypothesis, the exponent $K$  increases steadily from its non-interacting value $K=1$ at $\lambda=0$. It is impossible to follow it numerically beyond $\lambda \simeq 3$ because the floating phase becomes too narrow, but a linear extrapolation of the results beyond $\lambda=3$ suggests that it will reach the critical value $K=9/8$ around $\lambda=7$, hence that the floating phase has to terminate at a Lifshitz point located at the end of the PT line at $\lambda \simeq 7$. 

\begin{figure}[t!]
\centering 
\includegraphics[width=0.45\textwidth]{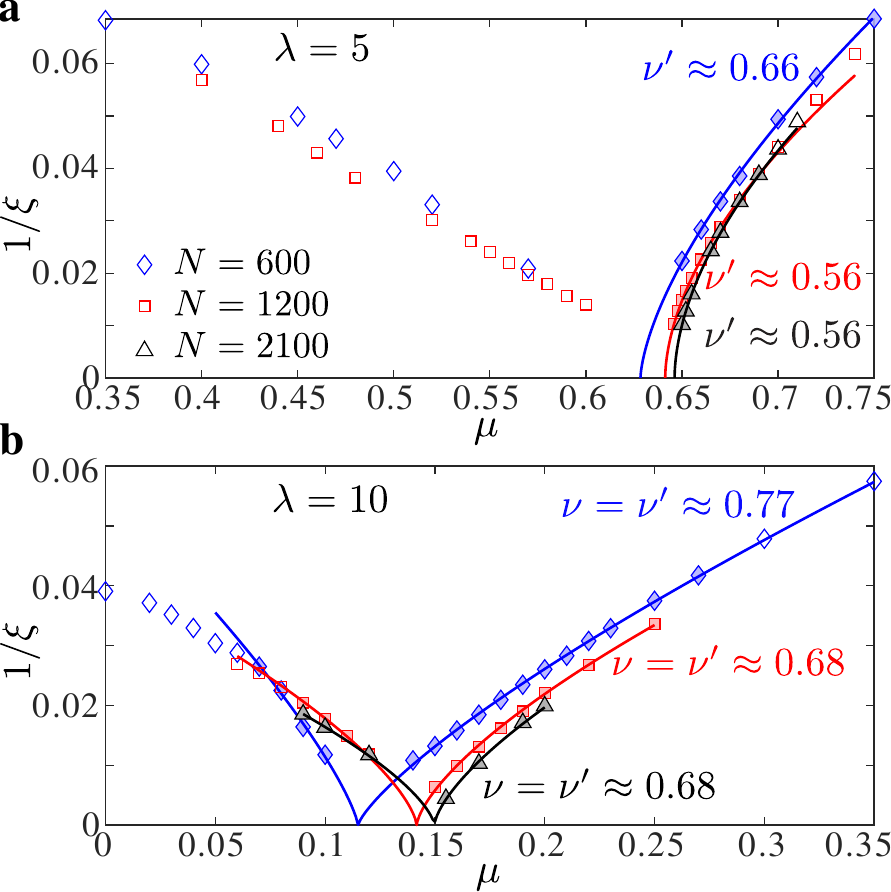}
\caption{Inverse of the correlation length on both sides of the Lifshitz point. {\bf a} $\lambda=5$, below the Lifshitz point. Solid lines are fits in the dirsordered phase with $|\mu_c -\mu|^{\nu^\prime}$. The value of the critical exponent ${\nu^\prime}$ decreases when the system size increases, and it is within 10\% of the Pokrovsky-Talapov prediction ${\nu^\prime}=1/2$. {\bf b} $\lambda=10$, above the Lifshitz point. Solid lines are fits of the data points on both sides of the transition with a unique critical transition point and equal critical exponents $\nu={\nu^\prime}$. The extracted values of the critical exponent remain well above $1/2$ and are in reasonable agreement with the critical exponents $\nu={\nu^\prime}\simeq 2/3$ reported previously in other studies of the chiral transition for the period-3 commensurate-incommensurate transition. The data points used for the fits are marked with filled symbols. }
\label{fig:cuts}
\end{figure}

As an independent check, we have kept track of the width $\Delta \mu$ of the floating phase as a function of $\lambda$ (see inset of Fig.\ref{fig:Kincrease}). A simple polynomial fit as a function of $1/\lambda$ suggests that the floating phase disappears at $1/\lambda\simeq 0.14$, in good agreement with $\lambda \simeq 7$.

To further check this prediction, we have carefully looked at the nature of the transition across two cuts that intersect the transition out of the disordered phase at $\lambda = 5$ and $\lambda = 10$ shown in Fig.\ref{fig:phasediag}. As shown in Fig.\ref{fig:cuts}, there is a clear difference in the way the correlation length diverges along these two cuts. For the lower cut, the correlation length is expected to diverge exponentially at the KT transition, algebraically with exponent $\nu^\prime=1/2$ at the PT transition, and to be infinite in between. In Fig.\ref{fig:cuts}{\bf a} one can clearly see the asymmetry between the left and right branches, signaling the existence of two different quantum phase transitions. 
% By fixing the location of the quantum critical point, we compute the effective critical exponent $\nu_\mathrm{eff}$ along the cut from pairs of consecutive data points. These results are presented in Fig.\ref{fig:cuts}(b). On the disordered side, the critical exponent for all system sizes approaches $\nu^\prime=1/2$ , in agreement with the PT transition. On the other side of the transition the effective critical exponent is significantly larger and increases with the system size consistent with an exponential divergence at a KT transition slightly below the PT one.

Above the Lifshitz point, the transition is expected to be a direct chiral transition in the Huse-Fisher\cite{HuseFisher} universality class. In this case the correlation length is expected to diverge on both sides of the transition with the same exponent $\nu=\nu^\prime$. Solid lines show the results of the fit of the data points on both sides of the transition with a single critical exponent and a unique critical point $\mu_c$. The extracted value of the critical exponent is consistent with $\nu=\nu^\prime \simeq 2/3$, in agreement with recent quantum field theory results\cite{PhysRevB.98.205118}, with numerical results on a classical model expected to have a transition in the same universality class\cite{nyckees2020identifying}, and with the exact result $\nu=\nu'=2/3$ derived \cite{albertini,Baxter1989} for an integrable version of the chiral Potts model\cite{auyang1987}, and extended by Cardy to a family of self-dual models\cite{Cardy1993}.

\section{Summary and perspectives}

To summarize, we have found a simple physical argument in favour of the presence of Lifshitz points at commensurate melting in 1D models of Rydberg atoms, and we have demonstrated that it rightly predicts the location of the Lifshitz point in a model of hard-core bosons with 3-site terms. The core of the argument relies on a simple property of the Pokrosvky-Talapov transition in systems where the number of particles is not conserved, namely that the Luttinger liquid exponent is not constant along this transition because it is not fixed by the density. This argument applies to Rydberg atoms, where excited states are created and annihilated by a laser with appropriate Rabi frequency, and it provides a solid physical basis to the results recently obtained on this problem. Interestingly enough, it probably also applies to the old problem of commensurate melting of surface phases in the context of which Huse and Fisher came up with the suggestion that the transition could be in a new, non-conformal universality class if the non-vanishing fugacity of the domain walls is properly taken into account.  Indeed, the role of the particles is played by the domain walls in these systems, and the fugacity controls the density of dislocations. So with a non-vanishing fugacity one can expect that the exponent of the floating phase will change along the Pokrovsky-Talapov transition line.

It will be very interesting to revisit the investigation of various models in quantum 1D and classical 2D physics along the lines of the present work. In particular, it would be interesting to try and measure the exponent of the critical phases in various models of commensurate melting along the Pokrovsky-Talapov line, and hopefully to locate accurately the Lifshitz point using the criterion that it reaches the critical value of the KT transition.

Beyond the Lifshitz point itself, the possibility to determine the extent of the floating phase on the basis of a numerical investigation of the Luttinger liquid exponent also opens new perspectives in the field. Indeed, locating the KT transition by looking at the correlation length is notoriously difficult because it diverges so fast that it exceeds the accessible system sizes long before the transition, and it is well known in the context of the XY model that to calculate the spin stiffness of the critical phase leads to much more accurate results. Work is in progress along these lines.

Finally, the present results provide a strong motivation to further investigate the properties of the chiral universality class, whose realization at commensurate melting is more likely than ever, but whose characteristics are still partly elusive.

\section{Methods}

\subsection{Extraction of the Luttinger liquid exponent}

\begin{figure}[t!]
\centering 
\includegraphics[width=0.45\textwidth]{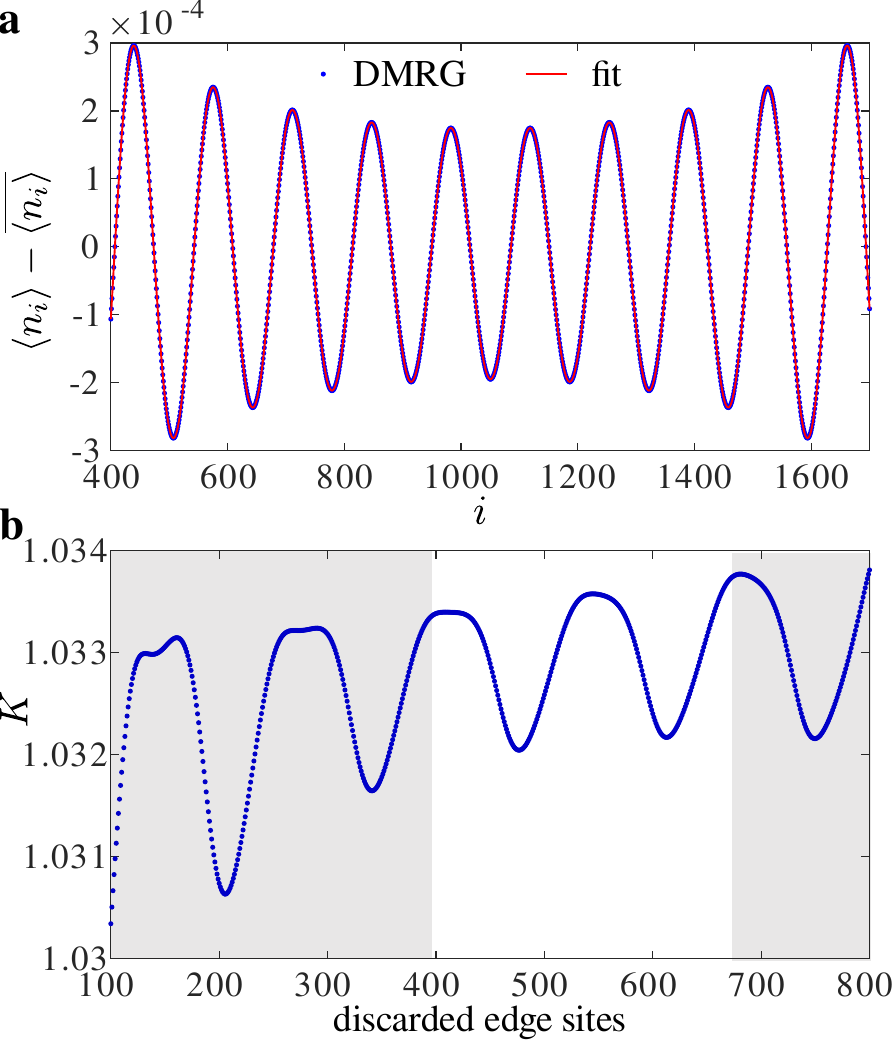}
\caption{(Color online) (a) Example of the local density profile obtained with DMRG for $N=2100$ sites at  $\lambda=1$ and $\mu=1.7945$ and centered around its mean value (blue dots). The red line is a fit to the CFT prediction $\propto \sin(qx +\phi_0)/\left[\sin(\pi x/(N+1))\right]^K$. The value of the incommensurate wave-vector extracted from this fit is $q\approx 0.01474688\pi$. The average error between the DMRG data and the fit is of the order of $10^{-8}$. To reduce finite-size effects we only fit the central part of the profile and discard a few hundreds sites at the edges. (b) Luttinger liquid exponent $K$ extracted from the fit as a function of the number of discarded sites at each edge. To get a better estimate of $K$ we take the average over an integer number of helices keeping a balance between the distance from the edges and the size of the middle domain (white region). }
\label{fig:fit}
\end{figure}

To extract the Luttinger liquid exponent inside the floating phase, we have fitted the Friedel oscillations of the local density profile induced by the open boundary conditions. An example of a typical fit is provided in Fig.\ref{fig:fit}. In the absence of incommensurability, boundary conformal field theory (CFT) predicts the profile to be of the form $\propto (-1)^x/\left[\sin(\pi x/(N+1))\right]^K$. To account for incommensurate correlations, we use the modified version $\propto \sin(qx +\phi_0)/\left[\sin(\pi x/(N+1))\right]^K$, which is expected to describe the asymptotic form of the scaling. Therefore, to reduce finite-size effects, we fit only the central part of the profile sufficiently far from the edges $x,(N-x)>>1$, as shown in Fig.\ref{fig:fit}(a). It turns out that the scaling dimension $K$ is sensitive to both the interval of the fit and the error in the wave-vector $q$. The latter problem is solved by increasing the accuracy of the fit to $\approx 10^{-8}$. The values obtained for $K$ are then averaged out over different numbers of discarded edge sites as shown in Fig.\ref{fig:fit}(b). To get a reliable fit, the central part used in the fit should contain sufficiently many helices. This prevents us from getting results in the immediate vicinity of the Pokrovsky-Talapov transition, where the wave-vector $q$ is very small, but, as we shall see, the exponent $K$ varies smoothly as a function of $\mu$, so it is still possible to get a precise estimate of $K$ in the vicinity of the PT line.

\begin{figure}[t!]
\centering 
\includegraphics[width=0.45\textwidth]{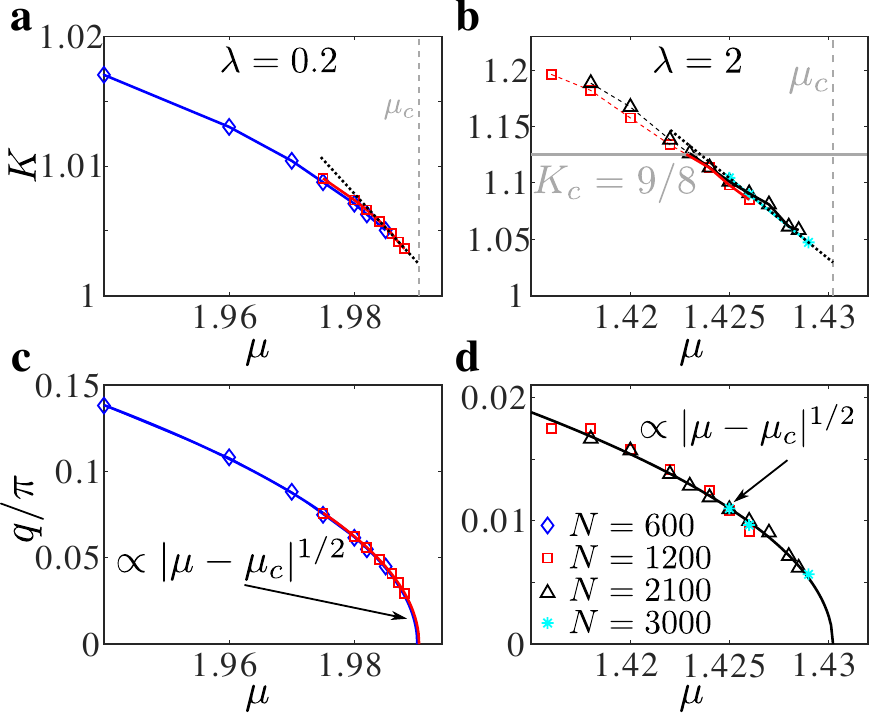}
\caption{Luttinger liquid parameter and incommensurate wave-vector for various cuts. {\bf a},{\bf b} Luttinger liquid parameter $K$ and {\bf c},{\bf d} incommensurate wave-vector $q$ as a function of the chemical potential $\mu$ for coupling constant $\lambda=0.2$ ({\bf a},{\bf c}) and  $\lambda=2$ ({\bf b},{\bf d}). Dotted lines are guides to the eye. The black dotted lines in panels {\bf a} and {\bf b} are linear fits of the 3-4 last points; in panels {\bf c} and {\bf d} we fit all available points to $|\mu_c -\mu|^{1/2}$. The error bars are smaller than the size of the symbols. Similar plots for $\lambda=1$ are presented in Fig.\ref{fig:BigFig_lambda1}{\bf a},{\bf b}}
\label{fig:LLatPT}
\end{figure}

We have extracted the Luttinger liquid parameter and the incommensurate wave-vector for various values of $\lambda$. In addition to  Fig.\ref{fig:BigFig_lambda1}{\bf a},{\bf b} for $\lambda=1$, we present our results for $\lambda=0.2$ and $\lambda=2$ in Fig.\ref{fig:LLatPT}. Note that for $\lambda=2$ the width of the floating phase is already very small, of the order of $7\cdot 10^{-3}$. The evolution of the Luttinger liquid parameter along the Pokrovsky-Talapov transition is presented in Fig.\ref{fig:Kincrease}.

\subsection{Density at the Pokrovsky-Talapov transition}

\begin{figure}[t!]
\centering 
\includegraphics[width=0.47\textwidth]{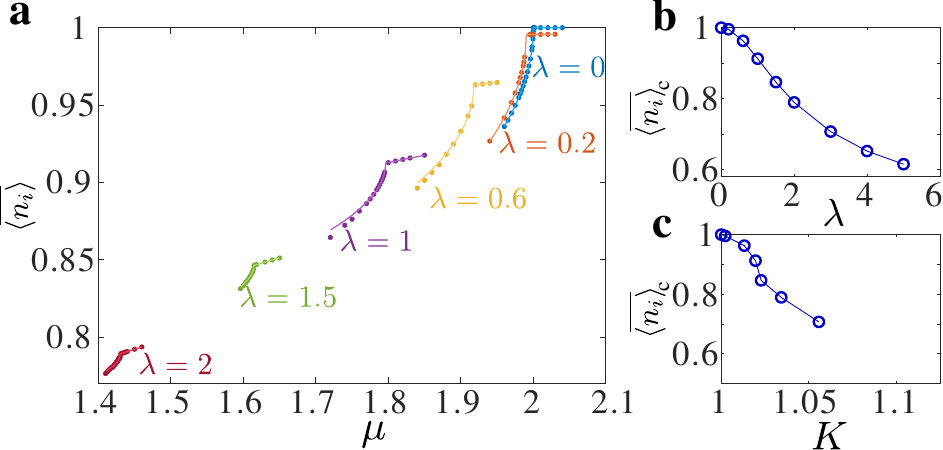}
\caption{Density at the Pokrovsky-Talapov transition. {\bf a} Average density across the Pokrovsky-Talapov transition for various fixed $\lambda$. Circles stand for the data points. The lines are the result of two types of fits: linear above the transition, and of the form $ a|\mu-\mu_c|^{1/2}+\langle n_i\rangle_c$ below it. {\bf b} Density along the Pokrovsky-Talapov line as a function of $\lambda$. The errors are smaller than the size of the symbols.  {\bf c} Density as a function of the Luttinger parameter $K$.}
\label{fig:dens}
\end{figure}

We argue that the Luttinger liquid exponent is not constant along this transition because it is not fixed by the density. In Fig.\ref{fig:dens} we summarize how the density evolves across the PT transition. We extract the density by averaging the local density $\langle n_i \rangle$. The interval over which we average always lies between two local maxima. This way, even if the wave-vector $q$ is close to zero, we obtain meaningful results. To reduce the edge effects we start with maxima located at 100-200 sites from the edges for $N=600$, and at $200-400$ sites for $N\geq 1200$. 

In order to find the density $\langle n_i\rangle_c$ at the PT transition  we fit the data point above the transition with a straight line and extract the value at the critical point determined from the incommensurate wave-vector $q$ as shown in Fig.\ref{fig:LLatPT}{\bf c},{\bf d}. Below the transition we fit the data with $ a|\mu-\mu_c|^{1/2}+\langle n_i\rangle_c$, where $a$ is the only fitting parameter.  As shown in panels {\bf b} and {\bf c} of Fig.\ref{fig:dens}, the density changes significantly along the PT line, and the change of density is correlated with the non-constant value of $K$ along this line.

%%%%%%%%%%%%%%%%%%%%%%%%%%%%%%%%%%%%%%%%%%%%%%

%%%%%%%%%%% 

\section{Acknowledgments}

We thank Titouan Dorthe, Matthjis Hogervorst, and Joao Penedones for useful discussions. This work has been supported by the Swiss National Science Foundation. The calculations have been performed using the facilities of the University of Amsterdam and the  facilities  of  the  Scientific  IT  and  Application Support Center of EPFL.

\bibliographystyle{ieeetr}
\bibliography{bibliography}

\begin{appendix}
 
\end{appendix}

\end{document}